\documentclass{article}

\usepackage{microtype}
\usepackage{graphicx}
\usepackage{subfigure}
\usepackage{booktabs} 

\usepackage[utf8]{inputenc} 
\usepackage[T1]{fontenc}    
\usepackage{hyperref}       
\usepackage{url}            
\usepackage{booktabs}       
\usepackage{amsfonts}       
\usepackage{nicefrac}       
\usepackage{microtype}      

\usepackage[final]{neurips_2020}

\usepackage{natbib}
\setcitestyle{numbers}

\usepackage{amsmath}
\usepackage{amsfonts}
\usepackage{amsthm}
\usepackage{mathtools}
\usepackage{amssymb}
\usepackage{complexity}
\usepackage{relsize}
\usepackage{graphicx, color}
\usepackage{epsfig}
\usepackage{caption}
\usepackage{multirow}
\usepackage{enumitem}
\usepackage{balance}
\usepackage{flushend}
\usepackage{epstopdf}
\usepackage{longtable}
\usepackage{stmaryrd}
\usepackage{pstricks}
\usepackage{pst-tree}
\usepackage{pst-all}
\usepackage{graphicx}
\usepackage{epstopdf}
\usepackage{lipsum}
\usepackage{xspace}
\usepackage{color}
\usepackage{framed}
\usepackage[normalem]{ulem}
\usepackage{float}
\usepackage{caption}
\usepackage{fancyvrb}
\usepackage{bm}
\usepackage[titletoc]{appendix}%

\renewcommand{\class}[1]{{\ensuremath{\mathsf{#1}}}}

\newcommand{\gen}{\ensuremath{\class{Gen}}\xspace}
\newcommand{\query}{\ensuremath{\class{Query}}\xspace}
\newcommand{\enc}{\ensuremath{\class{Enc}}\xspace}
\newcommand{\Enc}{\ensuremath{\class{Enc}}\xspace}
\newcommand{\Gen}{\ensuremath{\class{Gen}}\xspace}
\newcommand{\dec}{\ensuremath{\class{Dec}}\xspace}
\newcommand{\Dec}{\ensuremath{\class{Dec}}\xspace}
\newcommand{\eval}{\ensuremath{\class{Eval}}\xspace}

\newcommand{\vect}[1]{\mathbf{\ensuremath{#1}}}

\renewcommand{\k}{\ensuremath{K}}

\newcommand{\vq}{\vect{q}}

\newcommand{\ignore}[1]{}

\renewcommand{\A}{\mathcal{A}}
\renewcommand{\L}{\mathcal{L}}
\renewcommand{\S}{\mathcal{S}}

\renewcommand{\cc}[1]{\class{#1}}

\newcommand{\setup}{\class{Setup}}

\newtheorem{theorem}{Theorem}[section]

\newtheorem{definition}[theorem]{Definition}

\newcommand{\larr}{\ensuremath{\leftarrow}}
\newcommand{\rarr}{\ensuremath{\rightarrow}}

\renewcommand{\R}{\ensuremath{\mathcal{R}}}
\renewcommand{\D}{\ensuremath{\mathcal{D}}}

\newcommand{\T}{\ensuremath{\mathbf{T}}}

\renewcommand{\k}{\ensuremath{\sf K}}
\newcommand{\sk}{\ensuremath{\sf sk}}
\newcommand{\pk}{\ensuremath{\sf pk}}

\newcommand{\she}{\ensuremath{{\sf SHE}}}
\newcommand{\SHE}{\ensuremath{{\sf SHE}}}

\newcommand{\Real}{\ensuremath{\mathbf{Real}}}
\newcommand{\Ideal}{\ensuremath{\mathbf{Ideal}}}

\renewenvironment{proof}{\textbf{Proof:}}{\hfill$\blacksquare$}

\newcommand{\pprotocol}[5]{
{\begin{figure*}[#4]
\centering
\begin{center}
\advance\leftskip-1cm
\advance\rightskip-1cm
\setlength{\protowidth}{\textwidth}
\addtolength{\protowidth}{\intextsep}
\fbox{
        \hbox{\quad
        \begin{minipage}{\protowidth}
        #5
        \end{minipage}
        \quad}
        }
        \caption{\label{#3} #2}
\end{center}
\vspace{-5ex}
\end{figure*}
} }

\newlength{\protowidth} \protowidth \linewidth

\def\shownotes{1}

\mathchardef\mhyphen="2D
\ifnum\shownotes=1
\newcommand{\authnote}[2]{{\textcolor{red}{\textsf{#1 Notes: }\textcolor{blue}{ #2}}}}
\else
\newcommand{\authnote}[2]{}
\fi

\usepackage{url}
\usepackage{breakurl}

\newcommand{\OPE}{\cc{OPE}}
\newcommand{\xgboost}{\cc{XGBoost}}
\newcommand{\ppxgboost}{\cc{PPXGBoost}}
\newcommand{\encml}{\cc{EncML}}

\newcommand{\proxy}{\cc{Proxy}}

\usepackage{breakurl}

\title{Privacy-Preserving XGBoost Inference}

\author{%
    Xianrui Meng \ \ \ \ Joan Feigenbaum\thanks{Amazon Scholar and Grace Hopper Prof. of Computer Science at Yale University}\\
    AWS Cryptography Algorithm\\
  \texttt{\{xianru, jfbaum\}@amazon.com} \\
}

\begin{document}
\maketitle
\begin{abstract}
Although machine learning (ML) is widely used for predictive tasks, 
there are important scenarios
in which ML cannot be used or at least cannot achieve its full potential. 
A major barrier to adoption is the sensitive nature of 
predictive queries.  Individual users may lack sufficiently rich datasets
to train accurate models locally but also be unwilling to send sensitive
queries to commercial services that vend such models.
One central goal of {\it privacy-preserving machine learning} (PPML) is 
to enable users to submit {\it encrypted queries} to
a remote ML service, receive encrypted results, and decrypt them locally.
We aim at developing practical solutions for real-world privacy-preserving ML inference problems.
In this paper, we propose a privacy-preserving $\xgboost$ 
prediction algorithm, which we have implemented
and evaluated empirically on AWS SageMaker.  Experimental results indicate that
our algorithm is efficient enough to be used in real ML production environments.
\end{abstract}

\section{Introduction}\label{sec:intro}
Machine Learning (ML) plays an important role in daily life. 
Pervasive use of digital devices and networks produces massive amounts of 
data that are analyzed to reveal patterns and correlations that, in turn, are 
used to draw conclusions or make predictions. 
Diverse applications that make successful use of ML include market forecasting, 
service personalization, voice and facial recognition, autonomous driving, 
health diagnostics, education, and security analytics. 

Clearly in tension with the utility of ML is the desire of individuals and
organizations for data privacy.
Both the input and the output of an ML prediction may be highly personal,
confidential information and may be constrained by regulations.  For example,
students' confidential educational records are governed by FERPA\ --\ the
Family Educational Rights and Privacy Act. 
Organizations that produce valuable models may 
wish to sell access to them on a pay-per-prediction basis and must protect 
them as one would any valuable form of digital intellectual property. The 
need to maintain the privacy of data subjects, to protect intellectual 
property, and to keep commercially valuable instances and predictions 
confidential motivate the study of {\it privacy-preserving machine learning} 
(PPML). 

Gentry has shown that one can perform arbitrary 
computations on encrypted data using fully homomorphic-encryption 
(FHE)~\cite{G09}.  In principle, FHE could fully resolve the tension between
utility of ML and data-privacy requirements, but there are application 
scenarios in which it is prohibitively computationally expensive by orders of 
magnitude.  In these scenarios, it is natural to seek specialized 
homomorphic-encryption (SHE) schemes that are more efficient.  
We provide one such scheme in this paper.


{\it Extreme Gradient Boosting} ($\xgboost$)~\cite{DBLP:conf/kdd/ChenG16} is 
an optimized, distributed, gradient-boosting ML framework designed to be 
highly efficient, flexible, and portable. 
It performs {\it parallel tree boosting} that solves 
many classification and regression problems quickly and accurately. For example,
17 of the 29 challenge-winning solutions published on 
Kaggle’s blog in 2015 used $\xgboost$~\cite{KaggleComp},
Facebook uses it to predict click through on 
advertisements~\cite{conf/adkdd/HPJXBS/ADKDD14}, and it is very popular on
Amazon’s managed-cloud ML platform 
SageMaker.\footnote{\url{https://aws.amazon.com/sagemaker/}}

We present $\ppxgboost$,
a privacy-preserving $\xgboost$-prediction algorithm, in 
Section~\ref{sec:algorithm}.  In Section~\ref{sec:security}, 
we explain the security definition and the privacy properties that our algorithm achieves. Experimental
results are given in Section~\ref{sec:experiments}. Finally, we present open problems in 
Section~\ref{sec:open}.

\section{Privacy-preserving XGBoost}\label{sec:algorithm}

\subsection{Preliminaries}
\paragraph{XGBoost} Upon receiving a training dataset, the $\xgboost$ training algorithm produces 
an ML model in the form of a set $\{T_i\}_{i=1}^M$
of {\it classification and regression trees} (CARTs).  A CART is a 
generalization of a decision tree; while the latter produces a binary output,
thus classifying each input query as a ``yes'' or ``no'' instance of the 
phenomenon under study, a CART assigns to each input query a (real)
numerical score.  Interpretation of scores is 
application-dependent.\footnote{This is one of the roles of the 
hyper-parameter $\alpha$ referred to in Subsection~\ref{subsec:algorithm}} 
We use boldface lowercase letters, such as $\vect{v}$, to denote a vector of real numbers.
If $\vect{v}$ is an input query, and $\{y_i\larr T_i(\vect{v})\}_{i=1}^M$ is
the set of scores produced by the CARTs, then the final prediction ({\it
i.e.}, the overall score assigned to $\vect{v}$ by $\xgboost$) is typically 
$y = \sum_{i=1}^M y_i$.  Depending on the application, a $softmax$ function
may be applied to the $y_i$ to obtain the final prediction, but we restrict
attention to summation in this paper.
A full explanation of $\xgboost$ training and inference can be 
found in the original paper of Chen and Guestrin~\cite{DBLP:conf/kdd/ChenG16}. 


\paragraph{Homomorphic encryption}
Homomorphic encryption is a form of encryption that can perform arbitrary computation on plaintext values while manipulating only ciphertexts.  In this work, we use an additive SHE scheme. 
Specifically, let $\SHE=(\gen, \enc, \dec)$ be a public-key SHE that consists of three polynomial-time algorithms. $\gen$ is a probabilistic algorithm that takes a security parameter $k$ as input and returns a private and public key pair $(\pk, \sk)$. $\enc$ is a probabilistic encryption algorithm that takes as input a public key $\pk$ and
a message $m$ and outputs a ciphertext. $\dec$ is an algorithm that takes as input a private key $\sk$ and a ciphertext and returns the corresponding plaintext. 
In addition, $\SHE$ has an evaluation algorithm $\eval$ that supports any 
number of additions over the plaintexts: 
$\enc(\pk, (m_1 + \cdots + m_n)) = \eval\big(+, \enc(\pk,m_1), \ldots,
\enc(\pk,m_n)\big)$.

\paragraph{Order-preserving encryption} 
For $A, B \subseteq N$ with $|A| \le |B|$, a function $f :A \to B$ is 
{\it order-preserving} 
if, for all $i, j \in A$, $f(i) > f(j)$ if and only if $i > j$. 
We say that a symmetric encryption scheme $(\gen, \enc, \dec)$ with
plaintext and ciphertext spaces $\D$ and $\R$ is an 
{\it order-preserving encryption} ($\OPE$) scheme 
if $\Enc(\K, \cdot)$ is an order-preserving function  from $\D$ to $\R$, for all $K$ output by $\Gen(1^k)$. 
To make sense of the $>$ relation in this context, elements of $\D$ and $\R$ 
are encoded as binary strings, which are then interpreted as numbers. 

Throughout this paper, ``polynomial'' means ``polynomial in the security
parameter $k$.'' Formal definitions of these cryptographic concepts, including the Pseudorandom Function (PRF) family and semantic security can be found in \cite{KL08}. 

\pprotocol{s}{$\ppxgboost$: A privacy-preserving $\xgboost$ inference
algorithm}{fig:algo}{htbp}
{\small
\textbf{$\setup$ Phase:}\\
\textbf{Input}: Plaintext model $\Omega$; Security parameter $k$; Model hyper-parameter $\alpha$.

Proxy computes: 
    
    \begin{itemize}[noitemsep,nolistsep,label={-}]
        \item $K_1 \larr \OPE.\Gen(1^k)$;
        \item $(\pk, \sk) \larr \she.\Gen(1^k)$;
        \item Choose $f$ uniformly at random from ${\mathcal F}_k$ in the PRF family;
        \item Choose $K_2$ uniformly at random from $\{0,1\}^k$;
        \item For each CART $T_i\in \Omega$, construct $T_i^\prime$ in $\encml$ as follows:
        \begin{itemize}[noitemsep,nolistsep]
         \item $T_i^\prime$ is structurally isomorphic to $T_i$.  
              Let $\phi: T_i \rarr T_i^\prime$ be an isomorphism;
         \item For each internal node $x$ with value $\vect{v}$ in $T_i$, 
              assign the value $\OPE.\Enc(K_1, \vect{v})$ to $\phi(x)$;
         \item For each leaf $z$ with value $y_i$ in $T_i$, assign the value $\she.\Enc(\pk,$ $y_i)$ to $\phi(z)$;
         \item For each feature name $\ell$ used in $T_i$, create the corresponding feature pseudonym $\ell^\prime \larr f(K_2, \ell)$;
        \end{itemize}
    \end{itemize}

\textbf{Proxy sends to ML Module:} $\encml$; \ \ 
\textbf{Proxy sends to client:} $\alpha$, $K_1$, $K_2$, $f$, $\sk$;

\textbf{$\query$ Phase:}\\
\textbf{Input (to client):} Query $\vect{q}$. 
\begin{itemize}[noitemsep,nolistsep, label={-}]
\item Client computes:
    \begin{itemize}[noitemsep,nolistsep]
        \item For each feature name $l$ in $\vect{q}$, compute the corresponding feature pseudonym  
              $l'\larr f(K_2, l)$;
        \item Encrypt the plaintext value: $\vect{q^\prime} \larr \OPE.\Enc(K_1, \vect{q})$;
    \end{itemize}
\item \textbf{Client sends to ML Module:} $\vect{q'}$

\item ML Module computes:
    \begin{itemize}[noitemsep,nolistsep]
          \item For each $T_i^\prime \in \encml$, evaluate $T_i^\prime$ on $\vect{q^\prime}$ to obtain value $y_i^\prime$;
          \item Homomorphically sum the values: $y^\prime \larr \SHE.\eval(+, y_1^\prime, \ldots, y_n^\prime)$, where $n$ is the number of CARTs in $\encml$ ({\it note: this step is slightly different for computing softmax objective}) ;
    \end{itemize}      
    
    \item \textbf{ML Module sends to client:} $y^\prime$;
    \item Client decrypts the result: $y \larr \SHE.\Dec(\sk,$ $y^\prime$);
    \item Client interprets the result using the model hyper-parameter $\alpha$;
\end{itemize}      

}

\subsection{$\ppxgboost$ inference algorithm}\label{subsec:algorithm}


%

The $\ppxgboost$ algorithm is given in Figure~\ref{fig:algo}.
On the client side, there is an app
with which a user encrypts queries and decrypts results.  On the server
side, there is a module called $\proxy$ that runs in a trusted environment and
is responsible for set up ({\it i.e.}, creating, for each authorized user,
an encrypted model and a set of cryptographic keys)
and an ML module that evaluates the encrypted queries.

The inputs to the $\setup$ phase are an $\xgboost$ model $\Omega$, a model
hyper-parameter $\alpha$, and a security parameter $k$. During this phase, 
$\proxy$ generates, for each authorized user, the keys $K_1$ and $(\pk, \sk)$.
$K_1$ is the user's key for the (symmetric-key) OPE scheme. $(\pk, \sk)$ is the
user's key pair for the (public-key) SHE scheme.  
$\proxy$ then encrypts 
the node values in each CART $T_i$ in $\Omega$ to create an encrypted CART
$T_i^\prime$ in this user's encrypted model $\encml$.  For each internal
node in $T_i$ with value $\vect{x}$, the value of the corresponding node
in $T_i^\prime$ is $\OPE.\Enc(K_1,\vect{x})$. (Vectors of values are encrypted
and decrypted component-wise.) For each leaf in $T_i$ with value $y$,
the value of the corresponding leaf in $T_i^\prime$ is $\she.\Enc(\pk,$ $y)$.
Finally, the proxy sends $K_1$, $K_2$, $f$ and $\sk$ to the user's client and sends
$\encml$ to the ML module.

In the $\query$ phase, the client first encrypts its plaintext query $\vect{q}$
with OPE, {\it i.e.}, it computes $\vect{q^\prime} \larr 
\OPE.\Enc(K_1,\vect{q})$.  
It sends $\vect{q^\prime}$ to the ML module, 
which evaluates each $T_i^\prime$ in $\encml$ on input $\vect{q^\prime}$ 
to obtain a value $y_i$. The module computes 
$y^\prime \larr \she.\eval(+, y_1^\prime, \dots, y_n^\prime)$, where $n$ is the number of 
CARTs, and sends it to the client,
which decrypts to obtain the final result 
$y \larr \she.\dec(\sk,$ ${y^\prime})$. 

The correctness of this scheme follows directly from the properties of OPE
and SHE.  Because $a > b$ if and only if $\OPE.\Enc(K_1,a) > \OPE.\Enc(K_1,b)$,
for all $a$, $b$, and $K_1$, and the same $K_1$ is used to encrypt both
queries and internal-node values in $\Omega$, an encrypted query will travel
precisely the same path through each encrypted CART $T_i^\prime$ that the
corresponding plaintext query would have traveled through the 
corresponding plaintext CART $T_i$.  Because the leaf values $y_i$ 
in $\Omega$ have been encrypted using the additively homomorphic encryption
operation $\she.\Enc(\pk,$ $y_i)$, and $\sk$ is the decryption key that
corresponds to $\pk$, the plaintext $y$ corresponding to the
ciphertext sum $y^\prime$ is the sum of the individual plaintext 
values $y_i$ in leaves of $T_i$.

The proxy also chooses, for each authorized user, a function $f$ 
uniformly at random from ${\mathcal F}_k$ in the PRF family and a 
a key $K_2$ uniformly at random from $\{0,1\}^k$ for use with $f$.  This
function is used to generate pseudorandom ``feature names'' for vectors
of queries and node values. We defer discussion of this aspect of the 
algorithm until the full paper.

Note that the plaintext $(\Omega, \alpha, k)$ can be used by a very large user population, 
but a unique, personalized encrypted model must be created for each individual user.


\section{Privacy properties}\label{sec:security}

Ideally, we would like a privacy-preserving inference algorithm to hide all
information about the model, the queries, and the results from all adaptive
probabilistic polynomial-time (PPT) adversaries.  

For the $\setup$ phase, this means that the adversary 
should be able to choose a sequence $M_1$, $M_2$, $\ldots$, $M_n$ of plaintext 
models, submit them to an oracle, and receive the corresponding sequence 
$M_1^\prime$, $M_2^\prime$, $\ldots$, $M_n^\prime$ of encrypted models; it may
choose the sequence adaptively in the sense that its choice of $M_i$ may 
depend upon the oracle's answers $M_1^\prime$, $\ldots$, $M_{i-1}^\prime$.
After this adaptive, chosen-plaintext, oracle-query phase, the adversary is
presented with an encrypted model 
that it has not seen before, and it cannot infer
anything about the corresponding plaintext model.

For the $\query$ phase, this means that, for a fixed encrypted model 
$M^\prime$, the adversary should be able to choose a sequence $q_1$, $q_2$,
$\ldots$, $q_n$ of plaintext queries, submit them to an oracle, and receive
the corresponding sequence 
$\sigma = (q_1^\prime, r_1^\prime)$, $(q_2^\prime, r_2^\prime)$, 
$\ldots$, $(q_n^\prime, r_n^\prime)$ 
of encrypted queries and encrypted results; 
once again, it may choose $q_i$ based on $\sigma_{i-1} = (q_1^\prime,
r_1^\prime)$, $\ldots$, $(q_{i-1}^\prime, r_{i-1}^\prime)$.  After this
adaptive query phase, it cannot infer anything about the encrypted
model; furthermore, when subsequently presented with additional pairs 
$(q_{n+1}^\prime, r_{n+1}^\prime)$, ..., $(q_{n+j}^\prime, r_{n+j}^\prime)$,
it cannot infer anything about the corresponding plaintext
queries or answers. 

Known algorithms that achieve these ideal privacy properties
are not efficient enough for practical use.
As initiated in the work of Curtmola {\it et al.}~\cite{CGKO06} and
Chase and Kamara~\cite{CK10}, one can instead define acceptable {\it leakage
functions} and devise efficient algorithms that {\it provably} leak only 
the values of these functions.  In $\ppxgboost$, this
information may be leaked to the ML module and any party that observes the
inner workings of the ML module, the communication between $\proxy$ and the
ML module, or the communication between the client and the ML module. 

Because of space limitations we give the main ideas of our formal security definitions, $\ppxgboost$'s privacy properties, and our security proof in Appendix~\ref{sec:securityproof}.

\section{AWS SageMaker experiments}\label{sec:experiments}

\begin{table*}[thbp]
\begin{center}
\small
\begin{sc}
\begin{tabular}{lccccr}
\toprule
 & \multicolumn{2}{c}{time} & \multicolumn{2}{c}{Model size} \\
\midrule
Dataset & $\xgboost$ & $\ppxgboost$ & $\xgboost$ & $\ppxgboost$ \\
 \midrule
Amazon Synthetic Data  & $1ms$ & $0.43s$ 	 & 506KB & 4.2MB\\
Titanic 	&  $<1ms$ & $0.32s$ 	 & 3KB & 12KB \\
US Census  & $1ms$ & $0.49s$		 & 210KB & 2.5MB\\
\bottomrule
\end{tabular}
\end{sc}
\end{center}
\vskip -0.1in
\caption{\small$\ppxgboost$ Performance}\label{tab:ppxgboost}
\end{table*}
For our experiments on $\ppxgboost$, we implemented the cryptographic protocols
in {\tt python3}. We instantiated the PRF using $\cc{HMAC}$, and we used 
Paillier encryption~\cite{Paillier99} for our additive SHE and 
Boldyreva {\it et al.}'s scheme~\cite{BCN11} for our OPE.

An overview of our system architecture can be found in Appendix~\ref{sec:system}.
All of our experiments were run on AWS.
The $\setup$ phase is deployed in 
AWS Virtual Private Cloud~\footnote{\url{https://aws.amazon.com/vpc/}} environment. 
The inference procedure is run on SageMaker using an \cc{ml.t2.large} instance and the Amazon Elastic Container Service~\footnote{\url{https://aws.amazon.com/ecs/}}. 
Our experimental results are summarized in Table~\ref{tab:ppxgboost}. 
We ran $\ppxgboost$ on three different models derived from three different datasets. One data is synthetically generated based on Amazon's dataset. The other two datasets are public datasets. On average, $\ppxgboost$ inference is approximately $10^3$ times 
slower than the plaintext version of $\xgboost$. The size of encrypted 
models is between four and nine times larger than that of the plaintext models.
The inference time includes the network traffic time.
This performance is sufficient for many inference tasks currently done on
smart phones that must query a remote server.

\section{Open problems}\label{sec:open}
Our initial version of $\ppxgboost$ is still quite limited. 
Currently, we can support binary classifications and 
multiclass classification using the softmax objective. 
Future work includes support for more learning parameters in the 
privacy-preserving version. 
Moreover, in our algorithm, we leverage the order-preserving encryption scheme 
to support comparisons. Comparison on semantically encrypted data is 
computationally expensive, but we plan to investigate the use of secure 
multiparty computation for this purpose. In particular, we will explore the 
use of two non-colluding servers that execute secure comparison for each 
internal node in an encrypted CART.

\balance

\bibliography{xianrui}
\bibliographystyle{plain}

\smallskip

\appendix 

\begin{appendices}

\section{System Architecture}\label{sec:system}
As mentioned in the paper, we deploy $\ppxgboost$ using AWS infrastructure. 
We set up an Amazon VPC environment for deploying the inference prototype. 
Amazon VPC environment allows the model provider to have a logically isolated section of the AWS Cloud, therefore; the $\ppxgboost$ provider can have complete control over the virtual networking environment. The $\proxy$ service is deployed in a trusted environment, similarly to Amazon's Key Management Services (KMS).
The ML module is run on the Amazon SageMaker platform, a fully managed machine learning service. The security of SageMaker its own relies on the traditional AWS's security model~\footnote{see \url{https://docs.aws.amazon.com/sagemaker/latest/dg/data-protection.html}}, including AWS Identity and Access Management (IAM), Amazon Macie, etc.

\begin{figure}[th!bp]
\centering
    \includegraphics[width=.8\columnwidth]{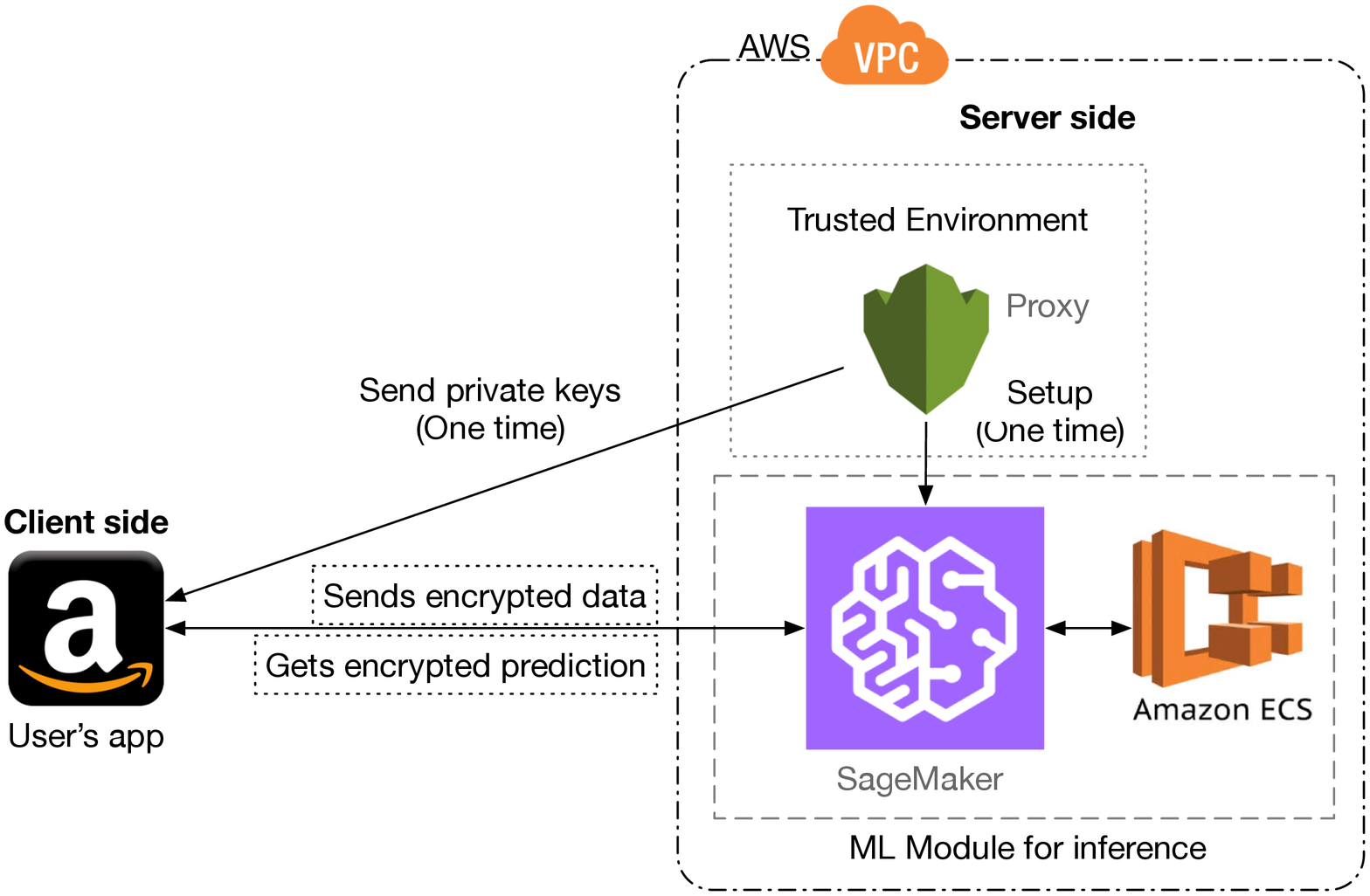}
    \caption{System architecture}\label{fig:sys}
\end{figure}

Amazon SageMaker makes extensive use of Docker containers for build and runtime tasks.
After the $\proxy$ produces an encrypted model $\encml$, we store $\encml$
to an S3 bucket (with proper permission configuration). 
We package our inference algorithm using Amazon's container service, Amazon ECS. 
When deploying the ML inference module, we upload the inference package in Amazon's
Elastic Container Repository and specify the encrypted model location in S3. We create an endpoint for this SageMaker instance to handle the encrypted queries.
After the client receives a private key, the client can send an encrypted query by querying a SageMaker endpoint.
The ML module computes the encrypted query and returns an encrypted result to 
the client.

\section{Privacy Definitions and Proof}\label{sec:securityproof}
At a high level, the security guarantee we require from privacy-preserving inference
scheme is that: $(1)$ given an encrypted ML model, no adversary can learn any
information about the model; and
$(2)$ given the view of a polynomial number of $\query$ executions for an
adaptively generated sequence of queries $\vq = (q_1, \ldots, q_n)$, no
adversary can learn any partial information about either the model or $\vq$.

Such a security notion can be difficult to achieve efficiently, so often one
allows for some form of leakage. Following \cite{CGKO06,CK10}, this is usually
formalized by parameterizing the security definition with leakage functions for
each operation of the scheme which in this case include the $\setup$ algorithm and
$\query$ protocol. 

\subsection{Security definition}\label{subsec:securitydef}

In our description of what it means for $\ppxgboost = (\setup, \query)$ 
to be secure, $\A$ is a semi-honest adversary, 
$\S$ is a simulator, $\L_\setup$ and $\L_\query$ are the leakage functions,
and $\sigma$ and $\sigma_{i-1}$ are as defined above.  The terms {\it Ideal}
and {\it Real} are used as they are in the literature on searchable 
encryption~\cite{CGKO06}. 

Let $\Omega$ and $\alpha$ be an $\xgboost$ model and hyper-parameter chosen 
by $\A$.  Let $v$ be a polynomially bounded function of $k$.
We consider the following two randomized experiments.  

\noindent {\bf $\Real(\Omega, 1^k, \alpha)$:}
\begin{itemize}[noitemsep,nolistsep]
\item Run the $\setup$ protocol:
$(\k, \encml) \larr \setup(Omega, 1^k, \alpha)$ 
\item Execute the $\query$ protocol $m = v(k)$ times.  In these executions,
Client (adaptively) chooses queries as well as performing its role in
Figure~\ref{fig:algo}; it chooses $q_1$ uniformly at random.
\item Output the sequence $\sigma = (q_1^\prime, r_1^\prime)$, $(q_2^\prime, 
r_2^\prime)$, $\ldots$, $(q_m^\prime, r_m^\prime)$ of encrypted queries and
results. 
\end{itemize}

\smallskip

\noindent{\bf $\Ideal_{\A, \S}(\L_\setup(\Omega), 1^k, \alpha)$:} 
\begin{itemize}[noitemsep,nolistsep]
\item Given $\L_\setup(\Omega)$, $1^k$, and $\alpha$, $\S$ generates 
an encrypted model $\encml$ and sends it to $\A$. 
\item $\A$ and $\S$ conduct $m=v(k)$ executions of the $\query$ protocol, 
in which $\S$ plays the role of Client by (adaptively) constructing a sequence 
$(q_1,\ldots,q_m)$ of queries, and $\A$ plays the role of the ML module.
$\S$ generates $q_1$ uniformly at random and uses 
$\L_\query(\Omega, \sigma_{i-1})$ to generate $q_i$, $2\leq i\leq m$. 
\item Output the sequence $\sigma = (q_1^\prime, r_1^\prime)$, $(q_2^\prime, 
r_2^\prime)$, $\ldots$, $(q_m^\prime, r_m^\prime)$ of encrypted queries and
results. 
\end{itemize}

\smallskip

The gist of our security definition is that a PPT observer cannot distinguish
between outputs of the $\Real$ experiment, which runs the protocol in 
Figure~\ref{fig:algo}, and the $\Ideal$ experiment, in which a simulator that
knows the values of the leakage functions plays the role of the client and the 
adversary plays the role of the ML module. A {\it distinguisher} $\D$ is an
algorithm that plays a refereed game with the adversary. 
In each round of the game, the referee obtains $\Omega$ and $\alpha$ from
$\A$, runs either $\Real(\Omega, 1^k, \alpha)$ or 
$\Ideal_{\A, \S}(\L_\setup(\Omega), 1^k, \alpha)$, and shows the output of
whichever experiment is run to the distinguisher.

Let $w$ be a polynomially bounded function of $k$.  
The entire {\it distinguishing game} proceeds as follows.  Fix a security
parameter $k$. For $w(k)$ rounds, $\D$ may ask the referee to run either
$\Real(\Omega, 1^k, \alpha)$ or 
$\Ideal_{\A, \S}(\L_\setup(\Omega), 1^k, \alpha)$ and show him the output.
The referee then chooses $b\in \{0,1\}$ uniformly at random; if $b=0$, it
runs $\Real(\Omega, 1^k, \alpha)$, and, if $b=1$, it runs 
$\Ideal_{\A, \S}(\L_\setup(\Omega), 1^k, \alpha)$.  It shows the output to
$\D$, and $\D$ outputs its best guess $b^\prime$ of the value of $b$.  The
{\it distinguisher's advantage} in this game, 
which is a function of the security parameter $k$, is 
$|\Pr\left[b=b^\prime\right] - \frac{1}{2}|$.

\begin{definition}\label{def:secureDef}
\rm
We say that $\ppxgboost$ is {\it adaptively} $(\L_\setup,\L_\query)$-{\it 
semantically secure} if, for every PPT adversary $\A$, there exists a 
PPT simulator $\S$ for which every PPT distinguisher's advantage is
negligible in $k$.
\end{definition}

\subsection{Leakage profile}\label{subsec:leakageprofile}

We now describe the leakage functions that specify, in the sense of
\cite{CGKO06,CK10}, the information that $\ppxgboost$ is willing to leak for
the sake of efficiency.

\textbf{Setup leakage.} 
Recall that the $\setup$ phase of $\ppxgboost$ takes as one of its
inputs a plaintext model $\Omega$ and gives as one of the outputs an 
encrypted model. The plaintext model consists of a set of CARTs.
Setup leakage in $\ppxgboost$ is a function $\L_\setup(\Omega)$ of the
plaintext model; it consists of 
the number of CARTs in $\Omega$, the depth of each CART, and, for 
each internal node $w$ in each CART,
which of $w$'s two children has the smaller value.  Note that the numerical
values of the nodes are {\it not} leaked; this is true of both internal
nodes and leaves.  In the high-level descriptions of $\setup$ given in
Section~\ref{sec:algorithm} and Figure~\ref{fig:algo}, 
the entire structure of each CART is leaked, but, 
in practice, it is straightforward to pad each CART out to a complete binary
tree of the appropriate depth without changing the results of the computation.

\textbf{Query leakage.}
During the $\query$ phase of $\ppxgboost$, the client and ML module exchange
a sequence $\sigma = (q_1^\prime, r_1^\prime)$, $(q_2^\prime, r_2^\prime)$, 
$\ldots$, $(q_n^\prime, r_n^\prime)$ of encrypted queries and encrypted results.
Query leakage in $\ppxgboost$ is a function 
$\L_\query(\Omega, \sigma)$ of the plaintext model and this sequence.
It consists of a {\it query pattern} and the {\it set of paths}
that are traversed during the execution of the encrypted queries. 
For every encrypted query $q^\prime$ in $\sigma$, this phase of 
$\ppxgboost$ leaks the number of times it appears in $\sigma$ and where it 
appears; that is, for every $q^\prime$, 
the query phase reveals the set of $i$, $1\leq i\leq n$, 
such that $q_i^\prime = q^\prime$.  In addition,
for each $q_i^\prime$ and each encrypted CART, the path from the root to a 
leaf in that CART that is traversed during the evaluation of $q_i^\prime$ is 
leaked to the ML module and to any party that can observe the inner workings 
of this module while queries are executed. Note that the query pattern and
set of paths is well defined for each prefix $\sigma_i$ of $\sigma$. 
Crucially, the decryptions of the queries and results are {\it not} leaked.

%

\subsection{Main idea and interpretation of the security 
proof}\label{subsec:proof}
To prove that the {\it only} information leaked by $\ppxgboost$ is 
$\L_\setup$ and $\L_\query$, we present a PPT algorithm 
that is given $1^k$, $\alpha$, $\L_\setup(\Omega)$, and 
$\L_\query(\Omega,\sigma)$ as input 
and {\it simulates} the behavior of $\ppxgboost$'s $\setup$
and $\query$ phases. 
We provide the main idea of the security proof here and defer the full proof
to an expanded version of this paper.

Given $\L_\setup(\Omega)$, a simulator $\S$ can construct
a set $\{\T_i\}$ of CARTs with the required ordering of internal nodes 
by sampling random values from the co-domain of 
OPE.  For the leaves, $\S$ assigns values chosen at random from 
the co-domain of SHE.  Given $\L_\query(\Omega, \sigma)$, 
$\S$ follows, for each encrypted query, the appropriate paths in
each $T_i$ that it constructed and returns the leaf value.
The security properties of the OPE and SHE schemes ensure that the final
predications are not revealed.


\section{Related work}\label{sec:related}
Practical attacks on supervised learning systems that result in leakage 
of sensitive information about training datasets, models, or 
hyper-parameters can be found in, 
{\it e.g.}~\cite{ModelInversion15,GAN17,MembershipInversion17}. 
Among proposals to mitigate those attacks, the majority focus 
on classification models, including 
decision trees~\cite{PPDM}, 
SVM classification~\cite{PPSVM}, 
linear regression~\cite{PPLIN1,PPLIN2,PPLIN3}, 
logistic regression~\cite{PPLOG}, 
and neural networks~\cite{nn1,ecg,nn2}. 
Recently, a fast-growing number of works 
(e.g.,~\cite{popa,PPRidge,PPRidge2,gilad,aono, define, opt1,opt2,opt3,opt4,opt5,CryptoNets,secureML,opt6,opt7}) 
have achieved strong security guarantees in this setting by providing 
concrete security definitions and provably secure protocols that use
multiple cryptographic-computation techniques~\cite{mpc-book}. 
Another research thread has focused on privacy-preserving federated learning 
(see, {\it e.g.},~\cite{Bonawitz17}), in which multiple mobile users update a 
global model by sharing aggregated updates to model parameters using a
privacy-preserving, client-server protocol. Recently, Liu 
{\it et al.}~\cite{DBLP:journals/corr/abs-1907-10218} proposed a 
privacy-preserving boosting method for training $\xgboost$ models in the 
federated-learning setting.
\end{appendices}

\end{document}